\documentstyle[prl,tighten,aps]{revtex}
\begin{document}
  \title{Neutrino condensates at center of galaxies as background for the MSW mechanism}
\author{
 S. Capozziello$^{a,b}$\thanks{E-mail: capozziello@sa.infn.it},
 G. Iovane$^c$\thanks{E-mail: iovane@diima.unisa.it},
 G. Lambiase$^{a,b}$\thanks{E-mail: lambiase@sa.infn.it} }
\address {$^a$Dipartimento di Fisica "E.R. Caianiello",
  Universit\'a di Salerno, 84081-Baronissi (Sa), Italy,}
\address{$^b$INFN Sez. di Napoli - Gruppo Collegato di Salerno, Italy.}
\address{$^c$Dipartimento di Ingegneria  e
  Matematica Applicata, Universit\'a di Salerno,  84084 - Fisciano (Sa),
  Italy.}
\date{\today}
\maketitle
\begin{abstract}
The possibility is explored that  neutrino condensates, possible
candidates for the explanation of very massive objects in galactic
centers, could act as background for the
Mikheyev-Smirnov-Wolfeinstein mechanism responsible of neutrino
oscillations. Assuming a simple neutrino star model with constant
density, the lower limit of the mass squared difference of
neutrino oscillations is inferred. Consequences on neutrino
asymmetry are discussed.
\end{abstract}

\vspace{2. mm}


\section{Introduction}
 The nature of matter condensates at galaxy centers \cite{oort}
is nowadays widely discussed since no current proposed model
seems fully able to fit the data. For example, recent
observations toward the Center of our Galaxy indicate that it
hosts a super-massive dark compact  object, Sagittarius A$^*$
(Sgr A$^*$) \cite{genzel1}, which is an extremely loud radio
source of estimated   mass $M\sim 2.61\times 10^6 M_{\odot}$ and
radius $R\sim 0.016$pc$\sim 30.6$lds \cite{ghez,genzel96} and
several evidence for the presence of similar objects have been
found in other galaxies, in quasars, and active galactic nuclei in
a mass-range $10^6\div 10^9 M_{\odot}$ .

The most natural candidates to explain such peculiar objects could
be either a {\it single} super-massive black hole or a very
compact cluster of stellar size black holes \cite{genzel96}. This
last case has been ruled out by stability criteria \cite{sanders}
which give maximal lifetimes of the order of $10^8$ years which
are much shorter than the estimated age of standard galaxies
\cite{maoz}. The first hypothesis is much more supported since
similar super-massive black holes allow to explain the central
dynamics of several galaxies as M87 \cite{ford,macchetto}, or
NGC4258 \cite{greenhill}. However, if Sgr A$^*$ were a
super-massive black hole, its luminosity should be more than
$10^{40}$erg/s, in contrast to observations which give a
bolometric luminosity of $10^{37}$erg/s (this is the so called
{\it blackness problem} or the {\it black hole on starvation}).
Besides, the most recent observations probe the gravitational
potential at a radius larger than $4\times 10^{4}$ Schwarzschild
radii of a black hole of mass $2.6\times 10^{6}M_{\odot}$
\cite{ghez}, so that unambiguous proof that the super-massive
object at the center of Galaxies is a black hole is still lacking.

Several alternative models have recently appeared in literature:
for example, in \cite{boson}, the hypothesis that the Galactic
Center could consist of a super-massive boson star is investigated
and some proposals have been done in order to seek a signature
like, for example, the Cerenkov effect \cite{cerenkov}.

Viollier {\it et al.} \cite{viollier} proposed that the dark
matter at the center of galaxies could be made by non-baryonic
matter ({\it massive neutrinos}) which interacts gravitationally
forming super-massive balls in which the degeneracy pressure of
fermions balances their self--gravity. Such neutrino balls could
be formed in the early epochs during a first--order gravitational
phase transition and their dynamics could be reconciled with some
adjustments to the Standard Model of Cosmology \cite{viollier}.
The neutrino mass required for fitting observational data by mean
of super-massive degenerate neutrino stars is 10keV$\leq
m_{\nu_\tau}\leq 25$keV, $m_{\nu_\tau}$ being the mass of $\tau$
neutrinos making up the stars. The gravitational effects of such
a structure on the stars orbiting around it can be investigated
by gravitational lensing \cite{salv}, while astrophysical
constraints and signatures can be investigated by next generation
of $X$-ray satellites as XEUS or Constellation-X \cite{lecce}. In
\cite{lecce}, a more physical model is discussed taking into
account a neutrino ball in gravitational equilibrium of a
semi-degenerate fermion gas. Density and pressure within the ball
are defined by adopting a formalism based on a distribution
function in phase space, which allows to consider neutrinos with
a degeneracy degree varying from the center to the border of the
system. Limiting cases are fully degenerate fermion systems and
the classical isothermal spheres well-known in literature.

In this paper, we take into account the fact that a  condensate of
massive neutrino could naturally act as the background for the
Mikheyev-Smirnov-Wolfenstein (MSW) mechanism which gives rise to
neutrino oscillations. In Sect.II, we outline the simplest
neutrino star model in order to define the environment where such
oscillations should happen. The MSW effect and the active-sterile
oscillations are discussed in Sect.III. Conclusions are drawn in
Sect.IV.

\section{The Neutrino Stars Model}
 In this section, we shortly recall the simplest model of heavy neutrino
condensates, bound by gravity (for details, see
\cite{viollier,leimgruber}). In the Thomas--Fermi model for
fermions, the Fermi energy $E_{F}$ and the gravitational
potential binding the system are related by (in natural units)
\begin{equation}  \label{n1}
\frac{k_{F}^{2}(r)}{2 m_{\nu}}-m_{\nu}\Phi(r)=E_{F}
=-m_{\nu}\Phi(r_{0})\,,
\end{equation}
where $\Phi(r)$ is the gravitational potential, $k_{F}$ is the
Fermi wave number and $\Phi(r_0)$ is a constant chosen to cancel
the gravitational potential for vanishing neutrino density. The
constant  $r_{0}$ is the estimated size of the halo of the ball.
If one takes into account a degenerate Fermi gas, one gets
$k_{F}(r)=\left(6\pi^{2}n_{\nu}(r)/g_{\nu}\right)^{1/3}$, where
$n_{\nu}(r)$ is the neutrino number density,  assumed being the
same for neutrinos and antineutrinos within the halo. The number
$g_{\nu}$ is the spin degeneracy factor. If in the center of the
neutrino condensate there is a baryonic star (which is
approximated as a point source), the gravitational potential will
obey a Poisson equation where neutrinos (and antineutrinos) are
the source terms,
\begin{equation}  \label{n4}
\triangle\Phi=-4\pi Gm_{\nu}n_{B}\,,
\end{equation}
where $ n_{B}=N/V$ is the neutrino background number density with
$N$, the number of neutrinos making up the condensate,
$N=M/m_{\nu}$, and $V=4\pi R^3/3$ its volume. We have
\begin{equation}\label{vol}
n_B=\frac{N}{V}=\frac{3M m_{\nu}}{4\pi R^3}\,.\end{equation}
Eq.(\ref{n4}) is valid everywhere except at the origin. In the
case of spherical symmetry,  the Poisson equation reduces to a
radial Lan\'e--Emden differential equation \cite{viollier}.

The general solution of (\ref{n4}), has scaling properties and it
is able to reproduce the observations \cite{viollier}. In fact, a
degenerate neutrino star of mass $M=2.6\times 10^{6}M_{\odot}$,
consisting of neutrinos with mass $m\geq 12.0$keV for $g_{\nu}=4$,
or $m\geq 14.3$keV for $g_{\nu}=2$, does not contradict the
observations. Considering a standard accretion disk, the data are
in agreement with the model if Sgr A$^*$ is a neutrino star with
radius $R=30.3$ ld ($\sim 10^5$ Schwarzschild radii) and mass
$M=2.6\times 10^{6}M_{\odot}$ with a luminosity $L\sim
10^{37}$erg/s. Similar results hold also for the dark object
($M\sim 3\times 10^{9}M_{\odot} $) inside the center of M87.

For our purposes, we assume a simple model where the neutrino
density is constant around a point-like baryonic mass  (e.g. a
baryonic star). Such a model has been also considered in
\cite{salv} where lensing experiments have been proposed for
probing the existence of neutrino balls, without invoking the
presence of super-massive black holes at the center of our Galaxy.
If $n_B =constant$,  it follows, from Eq. (\ref{n4}), that
$\Phi\sim r^2$, in agreement with the Newton theorem applied to a
spherical mass distribution. Such a situation is achieved
considering a Fermi gas at temperature $T=0$ \cite{salv}.

\section{MSW Effect and $\nu_\tau-\nu_s$ Oscillation}
 In the framework of Violler {\it et al.} model, it is natural
to investigate the possibility that neutrinos crossing the {\it
neutrino background} making up the supermassive condensate, could
oscillate owing to the mechanism equivalent to MSW effect
\cite{MSW}. To be more specific, assuming that $\tau$ neutrinos
are the constituent of neutrino balls, we explore the possibility
that a beam of $\tau$ neutrinos ($\nu_\tau$) propagating in the
$\nu_\tau$-background, could oscillate in sterile neutrinos
$\nu_s$ ($\nu_\tau\to \nu_s$) through a neutral current
interaction. The neutrino $\nu_s$ has the property that does not
undergo electroweak interactions.

According to the standard model, the $\nu-\nu$ scattering is
described by the effective Lagrangian density
\begin{equation}\label{lag}
  {\cal L}_{eff}=-\frac{G_F}{\sqrt{2}}\,[\bar{\nu}_\tau\gamma^\mu(1-\gamma^5)
  \nu_\tau][\bar{\nu}_B\gamma^\mu(1-\gamma^5)\nu_B]
\end{equation}
where $\nu_B$ indicates background neutrino, i.e. $\nu_B\equiv
\nu_\tau$. For simplicity in what follows we write $\nu$ instead
of $\nu_L\equiv (1-\gamma^5)\nu/2$. Neutrinos forming the
condensate are non relativistic so that the axial current
contribution is negligible, as well as the spatial components of
the 4-current since it is related to the average velocity of
$\nu_B$. Then, by considering the average over the background, the
0-component of the 4-current gives, in agreement with standard
procedure \cite{MSW},
 \begin{equation}\label{aver}
<\bar{\nu}_B\gamma^0\nu_B>=<\nu_B^\dagger \nu_B>\equiv n_B\,.
 \end{equation}
Eq. (\ref{aver}) adds an amounts of $\sqrt{2}G_Fn_B$ to the enegy
of neutrinos propagating within the neutrino star. Equation of
evolution of neutrinos is (for simplicity we consider only the
{\it radial} motion)
  \begin{equation}\label{evolution}
 i\frac{d}{d r}\left(\matrix{
                           \nu_\tau \cr
                           \nu_s \cr }\right)=
 \Lambda\left(\matrix{
                           \nu_\tau \cr
                           \nu_s \cr }\right)\,,
 \end{equation}
where the mixing matrix is defined as \cite{MSW}
 \begin{equation} \label{matrix}
 \Lambda\equiv \left(
 \matrix{-\displaystyle{\frac{\Delta m^2}{2E}}\cos2\theta+a
                  &   \displaystyle{\frac{\Delta m^2}{2E}}\sin 2\theta
                \cr
    \displaystyle{\frac{\Delta m^2}{2E}}\sin 2\theta
                 & \displaystyle{\frac{\Delta m^2}{2E}}\cos2\theta  \cr
                 }\right)\,,
 \end{equation}
where $a\equiv \sqrt{2}G_Fn_B$. The mass eigenstates $\vert\nu_1>$
and $\vert\nu_2>$ are determined by diagonalizing the mixing
matrix in (\ref{matrix}): One writes the mass eigenstates as a
superposition of flavor eigenstates
 \begin{eqnarray}\label{13}
\vert\nu_1>&=&\cos\tilde{\theta}\vert
\nu_\tau>-\sin\tilde{\theta}\vert \nu_s>\,{,} \\
 \vert\nu_2>&=&\sin\tilde{\theta}\vert \nu_\tau>+
 \cos\tilde{\theta}\vert \nu_s>\,{,}
 \nonumber
 \end{eqnarray}
where the mixing angle $\tilde{\theta}$ is defined as
 \begin{equation}\label{14}
 \tan 2\tilde{\theta}=\frac{\Delta m^2\sin 2 \theta}
 {\Delta m^2\cos 2\theta-2aE}\,{.}
 \end{equation}
The corresponding eigenvalues are
\begin{equation}\label{eigen}
  \lambda_{1,2}=\pm \frac{\Delta m^2}{4E}
  \sqrt{(\cos 2\theta -b)\cos 2\theta+\sin^2 2\theta}\,,
\end{equation}
being $b=2aE/\Delta m^2$. The resonance condition occurs when
\begin{equation}\label{resonance}
  \cos 2\theta=b\,.
\end{equation}
We remember that the neutrino density $n_B$ is assumed constant.
 From Eq.
(\ref{resonance}), and using Eq.(\ref{vol}) one infers the
relation
\begin{equation}\label{adiab}
 \Delta m^2 \geq 7\times 10^{-14}\left(\frac{30.6\mbox{ld}}{R}\right)^3
 \frac{E}{\mbox{GeV}}
 \frac{M}{M_\odot}
 \frac{\mbox{eV}}{m_{\nu_\tau}}\, \mbox{eV}^2\,.
\end{equation}
Neutrino stars are characterized by the parameters $R\sim 30.6$ld
and $M\sim 2.6\times 10^6M_\odot$, thus Eq. (\ref{adiab}) becomes
\begin{equation}\label{lower}
 \Delta m^2\geq 1.4\times 10^{-7}\frac{E}{\mbox{GeV}}
 \frac{\mbox{eV}}{m_{\nu_\tau}}
 \,\mbox{eV}^2\,.
\end{equation}
The result does depend on neutrino energy and the mass of the
background neutrinos ($m_{\nu_\tau}\sim 10$keV). From
(\ref{lower}) we derive the allowed solutions for $\Delta m^2$:
  $$
 E\sim \mbox{MeV}, \qquad \Delta m^2 \geq
 10^{-14}\mbox{eV}^2,
 $$
 $$
 E\sim \mbox{GeV}, \qquad \Delta m^2 \geq
 10^{-11}\mbox{eV}^2,
 $$
 $$
 E\sim \mbox{TeV}, \qquad \Delta m^2 \geq
 10^{-8}\mbox{eV}^2,
 $$
 which agree with the best fit for active-sterile neutrinos of $\tau$
 type, $\Delta m^2 \leq 10^{-6}$ obtained from
 the big bang nucleosynthesis predictions and in the absence of lepton asymmetry.
 The case of neutrinos with very high energy
 $$
 E\sim \mbox{PeV}, \qquad \Delta m^2 \geq
 10^{-5}\mbox{eV}^2,
 $$
 is excluded.

In presence of lepton symmetry, the above bound $\Delta m^2 \leq
10^{-6}$ becomes $\Delta m^2 \leq 1$, and the range of energy
$1\div 10^6$GeV is permissible in order that neutrino stars might
induce an appreciable MSW effect

These results have non trivial consequences in relation to
neutrino asymmetries in the early Universe, widely discussed in
Refs. \cite{foot1,foot}.

\section{Discussion and Conclusions}
In this paper we have investigated the active-sterile neutrino
oscillations for neutrinos propagating in a neutrino star which
acts as background matter. Taking into account the neutral current
weak interaction, we have studied the resonance  condition, in
analogy to MSW mechanism for solar neutrinos.

The basic results here obtained are two-fold. From one side, our
results show that the present data on neutrino oscillations do not
contradict a possible existence of such exotic super-massive
objects. Actually much more investigation on this results is
necessary. First of all, we should take into account more
realistic models where the density of the condensate changes with
the radius \cite{lecce}. Furthermore, one should  investigate the
$\nu_e : \nu_\mu : \nu_\tau$ ratio which becomes  $\nu_e : \nu_\mu
: \nu_\tau '$, where $\nu_\tau '$ are the $\tau$-neutrinos number
reduced (with respect to the expected ones) as a consequence of
the $\nu_\tau - \nu_s$ oscillation induced by the neutrino star
environment (for this topic, see also \cite{ahlu}). On the other
side, our results show that oscillations affected by {\it matter},
i.e. induced by neutrino stars, could have effects on the
generation of neutrino asymmetries in the Universe \cite{foot}.
The latter occur as a consequence of active-sterile neutrino
oscillations, which generate a discrepancy of neutrino and
antineutrino number density (such a subject is widely discussed in
Refs. \cite{foot}). The lepton number of a neutrino flavor $f$ is
defined as
\begin{equation}\label{leptonnumber}
  L_f=\frac{n_{\nu_f}-n_{{\bar \nu}_f} }{n_\gamma (T)}\,,
\end{equation}
where $n_{\nu_f}$ ($n_{{\bar \nu}_f}$) is the number density of
neutrinos (antineutrinos), and $n_\gamma$ one of photons at
temperature $T$. Constraints on $L_f$ are derived from the cosmic
microwaves background and from big bang nucleosynthesis. In the
case of neutrinos $\nu_\tau$, which are of interest in our case,
bounds on the lepton number are $\vert L_{\nu_\tau}\vert\leq 6.0$.
The creation of the lepton number determines a strong modification
on the upper limit of the mass squared difference value. In fact,
it goes from $\Delta m^2 \sim 10^{-6}$eV$^2$, coming from the big
bang nucleosynthesis predictions, to $\Delta m^2 \geq 1$eV$^2$
\cite{foot,buras}.

As final comment, it is worth to point out that if at the center
of galaxies a black hole is present, then no oscillation phenomena
could occur, unless a violation of the equivalence principle is
invoked \cite{gasperini,leung}. This is opposite to results
obtained in this paper, where $\nu_\tau$-background matter affects
oscillations giving a new scenario and a strong signature for
future investigations aimed to probe the (possible) existence of
such exotic objects, the neutrino balls, at the center of
galaxies.

\end{document}